# High Rate Applications of Micromegas and Prospects

I. Giomataris

*DAPNIA, CEA Saclay, 91191 Gif sur Yvette Cédex, France*

Micromegas is presently in use by many experiments providing enhanced performance when used in high particle environment. Compared with other solid-state detectors or conventional gas counters Micromegas is robust and radiation hard. Recent improvements in the design and fabrication of low-cost Micromegas have led to a significant step for readout of large volume detectors. A novel Cherenkov counter with Micromegas photodetector readout is proposed for on-line particle identification and trigger at high rates.

## 1. INTRODUCTION

The next generation of particle detectors and next investigations on high energy and nuclear physics experiments will continue to push the limits of our current technology. More than likely, the next generation of experiments will require much more precise and accurate detectors with the ability to obtain high precision under extreme irradiation conditions. To meet this demand the new gaseous detector Micromegas has been invented in 1996. The concept MicroMegas (MicroMEsh GAseous Structure), developed for tracking applications, has shown great promise for handling high data rates with a rather low-cost structure. The working principle is based on the double stage parallel plate avalanche chamber described in details in previous papers [1,2,3,4].

The use of a narrow gap (25-150 μm) is key element in the Micromegas operation leading to great performance in several areas: stability, relative immunity to defects in flatness, and excellent energy resolution [2,5]. The small amplification gap produces a narrow avalanche giving rise to excellent spatial and time resolution: 12 μm accuracy has been already reached while time resolutions in the sub nanosecond range have been measured by several experiments [6,7].

The high radiation resistance of the Micromegas has been verified with intense x-ray irradiation up to 30mC/mm$^2$ without any sign of aging. The total deposited charge corresponds to an equivalent fluence of minimum ionizing particles of >10$^{15}$/cm$^2$ [8]. Similar results have been reported by many experiments in various high-flux particle beams.

The fast collection of the ions (<100 ns) suppresses the space charge effect, which limits the gain in the common gaseous detectors, and therefore the amplification gain does not saturate up to particle fluxes of 10$^9$ mm$^{-2}$s$^{-1}$. Due to secondary effects, however, a drop in the maximum achievable gain, before breakdown, has been observed. Even though, at a x-ray flux 10$^7$ mm$^{-2}$s$^{-1}$, the gain is still high enough to allow full detection efficiency [9].

## 2. MICROMEGAS IN PARTICLE PHYSICS EXPERIMENTS

The COMPASS fixed target experiment at CERN has pioneered the use of large 40x40 cm2 Micromegas detectors for tracking close to the beam line with particle rates of 25 kHz/mm$^2$ [10,11,12]. All detectors operate smoothly with single-plane efficiencies greater than 97%, with a spatial resolution of 70 μm at a rate of 4x10$^7$/s. We should point out that no degradation of performance was observed in the COMPASS detectors after several years of operation with an accumulated charge of a 2 mC/cm$^2$.

The NA48 experiment at CERN is using Micromegas detectors for the charged kaon upstream spectrometer consisting of three stations of Micromegas coupled to a time projection chamber (TPC). Tracking of kaons, at rates exceeding several 10$^7$/s, is performed with a time resolution of 0.6 ns and a spatial resolution better than 100 μm [13]. Notice the beam diameter is a few cm and therefore the particle flux is exceeding 10$^6$ cm$^{-2}$s$^{-1}$. Under these extreme conditions the performance of the detector was stable over many years of operation and an average accumulated charge of 2 mC/cm$^2$. Kaons are tagged with a momentum resolution of 0.6%, which improves significantly the resolution on missing masses. A gap of 25 μm was also developed improving by a factor of three the speed of the detector. A thinner gap of 12 μm is under study; it will provide an unprecedented fast-ion collection time of 2 ns in Ne and < 1ns in He surpassing by an order of magnitude the silicon solid state detector.

With the combined COMPASS and NA48 results it has demonstrated the large-scale feasibility and reliability of the novel detector concept, and several years of running is a proof of robustness and resistance to high radiation levels.

There are many other applications of the Micromegas detector especially in the neutron detection domain. These include:

- High rate neutron beam profiler in the CERN n-TOF [14,15] facility and neutron imaging in OAKRIDGE[16].
- A high-resolution detector for thermal neutron tomography[17] .
- A detector with high time resolution for fast neutron detection in inertial confinement fusion experiments providing sub-nanosecond time resolution [18].
- A novel compact, sealed detector called Piccolo-Micromegas, has been designed to provide in-core measurements of the neutron flux at a nuclear reactor and to give an estimation of the neutron energy [19]. First results inside a TRIGA nuclear reactor have proven the ability of the new detector to survive in such very-aggressive environment. The use of four neutron converters will permit to measure on-line the neutron flux and energy spectrum of the reactor.

Despite the great interest of its excellent spatial (< 14 μm) and time resolution (< 1 ns) as well as the high rate behaviour, this detector can operate in the single electron counting capability [20,21]. The latter can be used, for instance, to track single photoelectrons in the photodetection mode or to reconstruct with high accuracy low energy recoils as it is needed for x-ray polarization measurements [22,23].

Micromegas is also in use or under development for low energy neutrino experiments including neutrino oscillations, neutrino magnetic moment, coherent neutrino scattering and searches on solar axions or dark matter WIMPs [24-28]. Reading-out large volume TPCs with highly segmented anode plane is a key point for many applications [45-48]

## 3. MICROMEGAS IN HIGH INTENSITY HADRON BEAMS AND IMPROVEMENTS

In the previous sections we have presented the high rate capability of Micromegas in photon beams. In the case of high-energy hadrons an additional effect limits the rate capability due to highly ionizing recoils induced by charged particles close to the amplification gap producing accidental discharges [29]. Their probability (10-6-10-7 per incident hadron) strongly depends on the on the type of the incident particle and the nature of the gas mixture.[30].

We must point out that discharges are not destructive but they must be reduced in order to obtain a safe operation. Adding an additional amplification, for instance a second parallel plate [31], or a GEM [32] or a Micromegas-like preamplification [33] structure the spark probability drops to a negligible level, allowing to sustain very-high rat

Recently new developments are undertaken to further improve the rate capability of the detector in the single amplification mode. A first promising one is employing a resistive foil on top of the anode plane, which quenches the spark formation leading to a limited streamer as shown in figure 1 [34]. Similar effect has been observed in other parallel plate structure [35].

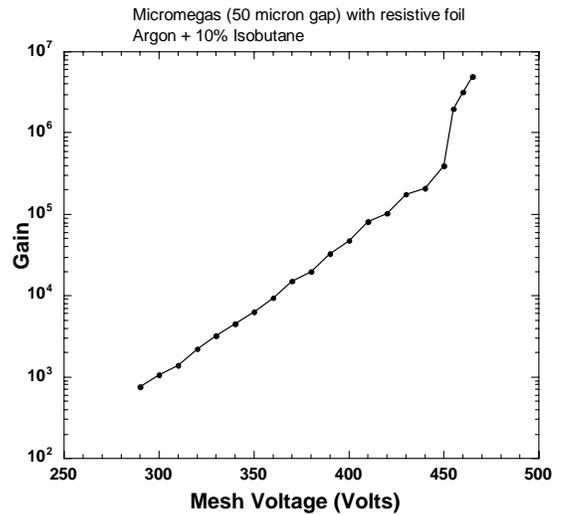

Figure 1: Gain versus the mesh voltage in Argon mixture using a Micromegas structure with a resistive foil on the anode(1.5 MOm surface resistance). We observe a linear behaviour up to a gain of $3x10^5$; beyond this value (the total charge in the avalanche is about $10^8$ e⁻ close to the Rather limit) the detector is reaching a limited streamer regime.

A second promising development is the segmentation of the grid which reduces both charge released during discharge process and the dead time during recovery. The grid segmentation is especially attractive in the case of the new technology 'Micromegas Bulk' that has been recently developed [36]. The basic idea is to build the whole detector in one process: the anode plane with the copper strips, a photo resistive film having the right thickness and the cloth mesh are laminated together at high temperature, forming a single object. By photolithographic method then the photo resistive material is etched producing the pillars. The new process allows easy implementation and provides light, low cost and robust detectors. Employing the Bulk technology large detectors have been built for the TPC prototype of T2K experiment. Using diamond tools or laser technology the cloth mesh was divided in many segments without loss of the detector performance as shown in Figure 2. The high voltage (V=300-400 Volts) is then distributed to the grid segments through individual resistors (R>100 kOm). This configuration divides the detector into several sub-detectors with capacitance roughly divided by the number of segments. Reducing the capacitance of each segment down to small values (C< 10 pF) is crucial for high rate operation :

- The total charge (Q= CV) released during discharging process does not exceed $2.5x10^{10}$ charges, a safe value for protection of front-end electronics.
- The spark process concerns only a specific segment, the entire detector area is not affected.
- The dead time (T=RC) of the specific segment is reasonably low of the order of T< 100 µs owing to a negligible efficiency loss.

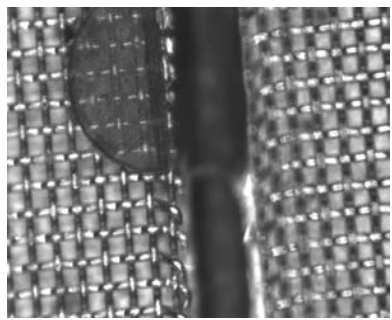 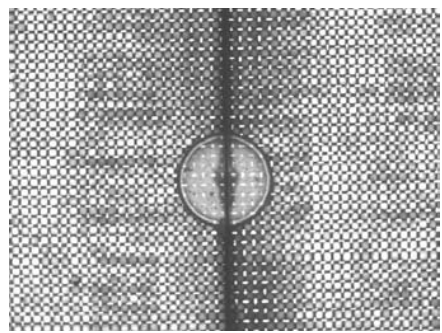

Figure 2: Cut of the woven mesh through the pillar (spacer 250 µm in diameter) with two different technologies: a) with diamond disk providing a slit thickness of about 120 µm b) with laser beam providing a slit thickness of about 20 µm

### 3.1. The Pion Blind Detector

The Hadron Blind Detector has been proposed by G. Charpak and myself in 1991[37]. The idea behind was the design of a novel threshold-Cherenkov counter capable to cope with substantial particle fluxes, for instance at the LHC collider; the aim was to resolve electron tracks inside complex hadronic events, with the advantage of being blind to most of hadrons. The new detector was based on a windowless concept enlarging the UV transparency of the radiator and allowing to reach a large quality factor [38,39].

An HBD prototype using a low-pressure $CF_4$ radiator and a 2 mm Parallel Plate Avalanche Chamber has been successfully tested in the SPS beam in 1992 [40]. A good efficiency for electrons and pion rejection has been obtained. For the first time a quality factor of $N_0$ =500 has been measured; this result is breaking the frontier of the conventional Cherenkov counter, limited by the absorption cut-off of the UV window that reduces the photon band width. It was the first demonstration of the windowless HBD concept.
A second important step forward was achieved by a SUNY-Stony Brook-MIT group [4]. Using an improved HBD, excellent efficiency and background rejection, has been measured.

A Hadron Blind Detector (HBD) is being developed for the PHENIX experiment at RHIC. The HBD will allow a precise measurement of electron-positron pairs from the decay of the light vector mesons and the low-mass pair continuum in heavy ion collisions. The detector consists of a 50 cm long radiator filled with pure $CF_4$ and directly coupled in a windowless configuration to a triple Gas Electron Multiplier (GEM) detector with a CsI photocathode evaporated on the top face of the first GEM foil [42,43].

An ideal candidate for improving the HBD detector is the use of reflective CSI-UV photocathodes evaporated on the MICROMEGAS mesh element [7]. Due to photon feed back suppression the detector is reaching high gains and is detecting single electrons with high efficiency and good time resolution. The use of the light Micromegas structure and a small amplification gap reduces the ionization produced by the charge particle as well as energy loss and delta-ray conversions; this a key element to reach an optimal signal to background ratio.

In this paper we introduce a new concept, called Pion Blind Detector, which is an evolution of the HBD detector, to address the following challenge:
In a new proposed experiment at CERN [44] the rare kaon decay $K^+ \to \pi^+ \nu \nu$ having very-low branching ratio << $10^{-10}$ will be studied. To reach this goal an increase of the kaon beam intensity by two order of magnitude, up to about 1 Ghz, is required: a real challenge for the first level trigger and on-line selection of charged kaons (only 5% of the total rate). A schematic view of the proposed detector is shown in figure 3. It consists of a gas radiator, a mixture of He and $CF_4$ to match the required refractive index and a spherical UV mirror reflecting the Cherenkov light produced by the beam particles (p60GeV/c). The reflected Cherenkov light is focused at the plane of the Micromegas UV photodetector. The refractive index is tuned in such way that the Cherenkov angle is saturated for pions and nearly saturated for kaons as shown in Figure2b. Under these conditions in the Micromegas projection plane two rings are detected; a large-saturated one corresponding to pions and a smaller to kaons. Signals coming from the inner kaon ring will give a clear and unbiased trigger for an efficient kaon on-line selection. Such fast signals (time resolution < 1 ns) are crucial for this experiment providing a first trigger and efficient particle identification.

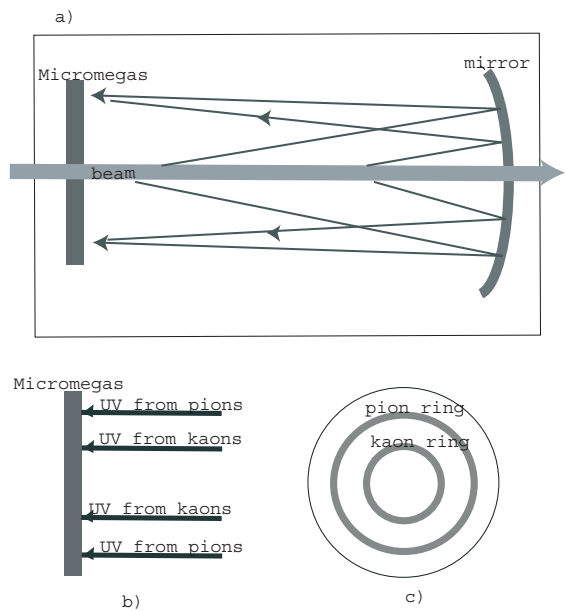

Figure 3: The principle of the Pion Blind Detector. a) Cherencov UV photons produced by the incident parallel particle beam are focused by a spherical mirror to the Micromegas photodetector plane, b) UV photons from pions and kaons are converted in different locations at the detector plane, c) Two rings are well separated, the larger diameter (saturated) is produced by pions while the internal ring by kaons.

## 4. SUMMARY

The Micromegas offers many advantages for high-rate high-flux environments. Large area detectors with excellent radiation resistance are easy to build with low material budget. Experience has also been gained in handling discharges initiated by heavy ionizing recoils induced by high-energy hadrons. Various schemes either in the single amplification or double amplification mode are giving an appropriate solution to this particular problem. An attractive issue is the use of a resistive foil on the anode plane quenching the discharge process or a segmented grid that reduces the discharge strength and dead time to a negligible level opening the way to operate at even higher event rates.

The novel Pion Blind Cherenkov counter concept could satisfy most of the requirements for very-fast particle identification and on-line trigger.